# Layer-by-layer disentanglement of Bloch states via frequency-domain photoemission

Woojoo Lee[1,†], Sebastian Fernandez-Mulligan[1,†,‡], Hengxin Tan[2], Chenhui Yan[1], Yingdong Guan[3], Seng Huat Lee[3,4], Ruobing Mei[3], Chaoxing Liu[3], Binghai Yan[2], Zhiqiang Mao[3], and Shuolong Yang[1,*]

[1]*Pritzker School of Molecular Engineering, The University of Chicago, Chicago, Illinois 60637, USA*

[2]*Department of Condensed Matter Physics, Weizmann Institute of Science, Rehovot 7610001, Israel*

[3]*Department of Physics, Pennsylvania State University, University Park, State College, Pennsylvania 16802, USA*

[4]*2D Crystal Consortium, Materials Research Institute, Pennsylvania State University, University Park, Pennsylvania 16802, USA*

[†]These authors contributed equally to the work.

[‡]Present affiliation: Department of History, Yale University, New Haven, Connecticut 06511, USA

[*]Corresponding author. Email: yangsl@uchicago.edu

**Abstract:** Layer-by-layer material engineering has enabled exotic quantum phenomena such as interfacial superconductivity and the quantum anomalous Hall effect. Meanwhile, deciphering electronic states layer-by-layer remains a fundamental scientific challenge. This is exemplified by the difficulty in understanding the layer origins of topological electronic states in magnetic topological insulators, which is key to understanding and controlling topological quantum phases. Here, we report a layer-encoded frequency-domain ARPES experiment on a magnetic topological insulator $(MnBi_2Te_4)(Bi_2Te_3)$ to characterize the layer origins of electronic states. Infrared laser excitations launch coherent lattice vibrations with the layer index encoded by the vibration frequency; photoemission spectroscopy tracks the electron dynamics, where the layer information is decoded in the frequency domain. This layer-frequency correspondence reveals a surprising wavefunction relocation of the topological surface state from the top magnetic layer into the buried second layer, reconciling the controversy over the vanishing broken-symmetry energy gap in $(MnBi_2Te_4)(Bi_2Te_3)$ and its related compounds. The layer-frequency correspondence can be harnessed to disentangle electronic states layer-by-layer in a broad class of van der Waals superlattices.

# Main

Manipulating electronic, magnetic, and lattice degrees of freedom layer-by-layer allows us to engineer materials' electronic properties at the level of single atomic sheets. This principle is manifested in the fabrication of magnetic topological insulators, where magnetism breaks the time-reversal symmetry and enables exotic topological phases of matter such as the axion insulators[1] and quantum anomalous Hall insulators[2]. Notably, the realization of these topological quantum phases relies on the fact that specific electronic states reside on the magnetic layers. A surgical probe of electronic states with both a milli-eV energy resolution and a layer-wise spatial resolution is key to unveiling the topological characters of interlaced quantum materials.

We focus on one of the most promising material candidates for realizing high-temperature topological orders – $(MnBi_2Te_4)(Bi_2Te_3)_n$ superlattices[3]. These superlattices consist of magnetic $MnBi_2Te_4$ (MBT) layers interleaved with non-magnetic $Bi_2Te_3$ (BT) layers. Varying the stacking order leads to a multitude of topological phases[1,2]. However, significant controversies have arisen over the coupling between the topological surface state (TSS) and the broken-symmetry MBT layers[3–12]. Angle-resolved photoemission spectroscopy (ARPES)[6] and scanning tunneling spectroscopy (STS)[13] studies have led to the theoretical proposal that the TSS can be decoupled from the top MBT layer. However, due to the lack of a direct layer resolution, the picture of TSS relocation remains debated. This is reflected from the alternative interpretation of the STS results[13] using Rashba-split electronic states[11], underscoring the pressing need to develop spectroscopic techniques with layer resolution to reveal the layer origins of the TSS and other key electronic states.

Here we demonstrate a layer-encoded frequency-domain ARPES experiment to disentangle electronic states layer-by-layer in (MBT)(BT) superlattices. Infrared pump pulses launch coherent phonon modulations, serving as the intrinsic tuning knob to probe electron-phonon coupling[14–16]. The crystal structure of our targeted material (MBT)(BT) differs layer-by-layer, yielding layer-specific coherent phonon frequencies. Subsequently, the induced energy oscillations of the electronic states are measured by time-resolved ARPES (trARPES), and linked to different layers through the decomposition in the frequency domain. This layer-frequency correspondence reveals a surprising wavefunction relocation of the TSS into the buried non-magnetic layers, and provides a critical hint to resolve the outstanding mystery of the near-gapless TSS. Our work is a milestone in ultrafast spectroscopy connecting the spatial and frequency domains, conceptualized as "seeing-by-listening".

We perform static micro ($\mu$)-ARPES measurements[17] to characterize the occupied electronic band structures on the two terminations of (MBT)(BT) (Fig. 1, **b** to **e**). On the BT termination, three parabolic bands near $\bar{\Gamma}$ are observed and labeled as α, β, and γ. On the MBT termination, only the α and β bands are identified. We characterize the α, β and γ bands as bulk conduction bands based on the comparison with density functional theory (DFT) (Supplementary Fig. S1) and previous

ARPES studies[10]. In addition, we observe the TSS on the BT termination which hybridizes with a bulk valence band (Fig. 1b and 1d). The Dirac point is observed near -0.37 eV. On the MBT termination, we observe a distinct electron-like pocket with a Fermi momentum of 0.19 Å$^{-1}$ (Fig. 1c and 1e). Circular dichroism ARPES (CD-ARPES) measurements on a series of (MBT)(BT)$_n$ compounds suggest that this feature likely originates from a pair of Rashba-split pockets with a coupling constant $\alpha_R \sim 0.7$ eV·Å (Supplementary Note 1). Therefore, we denote this pair of bands as Rashba-split states (RS). A V-shaped feature on the MBT termination with a binding energy of -0.2 eV originates from the TSS-RS hybridization (Supplementary Note 1).

We reveal the ultrafast electron dynamics on the BT termination by trARPES (Fig. 2). All investigations were performed consistently on the targeted sample position after ensuring a precise beam alignment between static- and tr-ARPES setups (Supplementary Note 2). To account for model-dependent systematic uncertainties, we fit time-dependent energy distribution curves (EDCs) to various models involving single or multiple Lorentzian peaks (Supplementary Note 3), and extract the consistent binding energy dynamics of the TSS (Fig. 2b). Fast Fourier Transforms (FFTs) of the coherent energy oscillations ($\delta E(t)$) exhibit a dominant 1.8 THz mode along with two sub-dominant modes at 1.2 and 3.4 THz (Supplementary Fig. S8). Comparison with Raman spectroscopy allows us to identify the 1.2 and 1.8 THz modes as the $c$-axis A$_{1g}$ modes associated with the MBT and BT layers, respectively (Fig. 2f)[18]. These frequencies are hardly changed in the superlattice family (MBT)(BT)$_n$ for $n \geq 1$, reflecting the fact that the 1.2 and 1.8 THz modes are intrinsic to the individual MBT and BT layers. Moreover, since the electronic structures of (MBT)(BT) near the Brillouin zone center are mainly contributed by the $p_z$ orbitals, the $c$-axis A$_{1g}$ modes lead to the strongest modulation of the orbital overlaps and thus the electronic binding energies. Fitted with cosine functions, the coherent responses to the 1.8 and 1.2 THz modes exhibit initial phases close to integer multiples of $\pi$ (Fig. 2c insets), manifesting the displacive excitation of coherent phonons (DECP)[19]. Notably, DECP leads to fully symmetric A$_1$ modes in the superlattice unit cell spanning one MBT layer and one BT layer[19]. This constrained symmetry is fully consistent with our mode assignments. Along the TSS dispersion, the existence and absence of the amplitude sign reversal for the MBT and BT A$_{1g}$ modes, respectively, both agree with theoretical calculations (Supplementary Fig. S9). By performing a pixel-wise Fourier transform of the trARPES data, we obtain the frequency-domain (FD) ARPES[14,16] maps at 1.8 and 1.2 THz (Fig. 2d and 2e). Notably, the oscillating TSS dispersion results in a pair of bands in FD-ARPES maps with opposite oscillation phases[20]. The momentum-independent and dependent phases along the TSS dispersion at 1.8 and 1.2 THz, respectively, well corroborate the EDC analysis in Fig. 2c. Lastly, the 3.4 THz mode can be attributed to either a second MBT A$_{1g}$ mode[18], or a putative surface symmetry-breaking BT mode[21]. This uncertainty precludes the use of the 3.4 THz mode to distinguish layer origins of electronic states.

Importantly, given the heterointerface of the (MBT)(BT) superlattice, there are extra phonon modes involving coordinated atomic movements from both the MBT and BT layers[18]. Among the extra phonon modes, we restrict our consideration to all the $A_{1g}$ modes within the enlarged unit cell. This is based on the fact that the 1.5 eV optical excitation with a penetration depth of ~16.7 nm[22] is largely homogeneous for the top few unit cells, and hence the symmetry selection rules imposed by the DECP mechanism can only allow fully symmetric modes.[19] In the frequency range of our interest, only the 1.2 THz MBT $A_{1g}$ mode and the 1.8 THz BT $A_{1g}$ mode are relevant to our observations, enabling us to use layer-specific phonon modes to probe the layer origins of electronic states. We also considered broken-symmetry modes at the material surface[21,23]. Yet, to the best of our knowledge no such modes have been identified in (MBT)(BT) in the vicinity of 1.2 THz and 1.8 THz.

The dominant coupling between the TSS and the BT $A_{1g}$ mode reveals a strong localization of the TSS on the top BT layer. This result is fully consistent with theoretical expectations, and sets the foundation to tackle the more nontrivial case of the MBT termination.

We turn to the trARPES measurement on the MBT termination. Importantly, the TSS and RS hybridize on this termination, resulting in a distinctively large electron-like pocket. This electron pocket has a stronger TSS character near $\bar{\Gamma}$, and a stronger RS character off $\bar{\Gamma}$ (Fig. 3**b**). We implement the aforementioned EDC analysis on this hybridized electron pocket and extract band energy dynamics. The obtained coherent oscillations suggest the existence of multiple frequency components (Fig. 3**c**). As evidenced by FFT (Supplementary Fig. S8**b**), two coherent modes exist at 1.8 and 1.2 THz, which are dominant near and away from $\bar{\Gamma}$, respectively. We fit the coherent energy oscillations with two-cosine functions, and extract the oscillation amplitudes at different momenta (Fig. 3**d**) which fully corroborate the FFT results.

We highlight the most substantial discovery on the MBT termination: the TSS and RS band characters are one-to-one mapped to the oscillation amplitudes of the BT and MBT $A_{1g}$ modes, respectively. This is further demonstrated by the FD-ARPES data for the 1.2 and 1.8 THz modes on the hybridized electron-like band (Fig. 3**e** and 3**f**). We will elucidate below that this observation suggests a wavefunction relocation of the TSS from the top MBT layer to the buried BT layer.

We invoke the fundamental theory of deformation potentials. Under the quasi-static assumption, lattice strain $S$ induces the energy shift $\delta\epsilon$ of a Bloch state at momentum $\boldsymbol{k}$ and band index $n$. The energy shift is expressed as[24],

$$\delta\epsilon_{\boldsymbol{k},n} = D_{\mu\nu}(\boldsymbol{k},n)S_{\mu\nu} \tag{1}$$
$$D_{\mu\nu}(\boldsymbol{k},n) = \langle \boldsymbol{k},n|\delta V^{\mu\nu}(\boldsymbol{r})|\boldsymbol{k},n\rangle$$

where $D_{\mu\nu}(\boldsymbol{k},n)$ is the deformation potential and $\delta V^{\mu\nu}(\boldsymbol{r})$ is the effective one-electron potential. Therefore, a suppressed $\delta\epsilon$ can be traced down to a vanishing $S_{\mu\nu}$ or $D_{\mu\nu}$. Notably, a vanishing $S_{\mu\nu}$ can occur for low-symmetry modes due to the short-lived driving force[25]. However, the coherent modes under consideration are fully-symmetric $A_{1g}$ modes, rendering the first scenario unlikely. Second, $D_{\mu\nu}$ can vanish when the overall parity of $\langle \boldsymbol{k}, n|\delta V^{\mu\nu}(\boldsymbol{r})|\boldsymbol{k}, n\rangle$ is odd[26]. However, our DFT calculations have predicted a non-vanishing $D_{\mu\nu}$ for the TSS energy shift in response to the MBT $A_{1g}$ mode, assuming the TSS is localized on the top MBT layer (Supplementary Fig. S10). The DFT calculations also demonstrate that other possible contributing factors such as electronic screening[27] and kinematic constraints[28] do not play major roles.

It is thus imperative to consider another scenario: the TSS wavefunction on the MBT termination is relocated into the BT layer (Fig. 4). The buried TSS will consequently exhibit a stronger coupling to the BT $A_{1g}$ mode instead of the MBT $A_{1g}$ mode, precisely reflecting our experimental observations. Microscopically, an increased van der Waals (vdW) gap between the top MBT layer and the second BT layer can push the TSS wavefunction into the deeper layers[13,29]. Our slab calculations demonstrate a substantial TSS relocation when the surface vdW gap exceeds the bulk counterpart by 10% (Supplementary Fig. S11 and S12), yet the increased vdW gap alone does not yield the RS. On the other hand, previous calculations incorporating Mn-Bi antisite defects yielded new band structures mimicking our observed RS[30]. Therefore, it is likely the combination of vdW gap increase and antisite defects that lead to the redistribution of the TSS wavefunction and the creation of the RS.

The layer origins of the TSS and RS shed light on the outstanding mystery of the near-gapless Dirac cones in antiferromagnetic (MBT)(BT)$_n$ superlattices[3–6,9–12]. An intense debate in the literature[3–12] centered around why the broken-symmetry gap on the MBT termination can vanish, despite the strong $c$-axis magnetic moments confirmed by magnetic force microscopy[31,32]. Our results suggest that the TSS on the MBT termination is partially pushed into the buried BT layer, and is less influenced by the surface magnetism. Such a wavefunction relocation has been discussed as a framework to reconcile experimental observations in ARPES[29] and STM[13] studies, yet this theoretical framework remains inconclusive due the lack of a direct layer resolution, leaving althernative interpretations possible. For instance, though previous STS measurements[13] used this framework to explain the observed anomalous quasi-particle interference pattern, the alternative scenario using Rashba-split states[11] challenged their interpretation. On the other hand, the experimentally evidenced wavefunction relocation in our work fully accounts for the near-gapless Dirac cone in (MBT)(BT) (Supplementary Fig. S12), and provides a hint to understand the phenomenology in pure MnBi$_2$Te$_4$ (Supplementary Note 4). Moreover, localization of the RS on the top MBT layer of (MBT)(BT) is directly revealed for the first time, explaining the energy gaps on the Rashba-like bands as observed in (MBT)(BT)$_n$ compounds[5,6,8,9]. To this end, our work resolves outstanding controversies [13,29,33] over the issue of broken-symmetry gaps in (MBT)(BT)$_n$ superlattices.

Our layer-encoded frequency-domain ARPES experiment can be generalized to all magnetic topological superlattices (MBT)(BT)$_n$ and beyond. This approach will provide the foundational physics insight for how to engineer digital superlattices to manipulate topology and other collective quantum phenomena. Importantly, layer-encoded frequency-domain ARPES requires the superlattice system to exhibit coherent phonon modes which are localized in individual layers. One needs to carefully examine the mode origins through the comparison with theory and with Raman spectroscopy. Symmetry-breaking modes can also exist on material surfaces or interfaces, which can provide additional information on the origins of electronic states.[21,23] Our experiment manifests a profound measurement philosophy of "seeing by listening," where the frequency information from each layer allows us to visualize the layer origins of electronic states. This new philosophy is particularly important and timely, as emergent many-body physics such as superconductivity and nontrivial topology have been explored in stacked and twisted superlattices[34–38], yet little is known about their layer-by-layer wavefunction distribution using currently available spectroscopies. Our work potentially fills this technological gap. To this end, we believe that the impact of this experiment goes beyond resolving the long-standing puzzles in $MnBi_{2n}Te_{3n+1}$, and establishes a new paradigm for probing the spatial distribution of wavefunctions.

## Methods

Sample growth

The MnBi$_4$Te$_7$ samples were synthesized using the melt growth method[39]. High purity Mn, Bi, and Te mixture were sealed in carbon coated quartz tubes under a high vacuum and heated in muffle furnace at 900 °C for 5 hours, followed by a cool-down to 595 °C in 20 hours. The mixture was then slowly cooled down to 585 °C in three days (~ 0.14°C/h) and annealed for one day. The single crystals were then obtained by water quenching at 585 °C. The samples were cleaved *in situ* under a base pressure lower than $8 \times 10^{-11}$ mbar at 15 K.

ARPES measurements

All ARPES measurements were carried out at 15 K on the Multi-Resolution Photoemission Spectroscopy (MRPES) platform established at the University of Chicago[17]. Ultrahigh resolution μARPES measurements were performed with a spatial resolution <10 μm and an energy resolution < 4 meV using 205 nm probe pulses with 80 MHz repetition rate. Ultrafast trARPES measurements were performed with an energy resolution of 17 meV and a time resolution of 115 fs at 200 kHz repetition rate. Pump and probe wavelengths were 800 nm and 206 nm, respectively. The incident fluence was 350 μJ/cm$^2$. The integration of high-energy-resolution and high-time-resolution light sources in one setup allowed us to combine complementary probes and obtain a holistic picture.

Computational details

All electronic structures were calculated with Density Functional Theory as implemented in Vienna ab-initio Simulation Package[40,41]. The electron-electron interaction was mimicked with the projected augmented wave method[42] with the exchange-correlation interaction approximated via generalized gradient approximation (PBE)[43]. A Hubbard U value of 5 eV was used for the *d* electrons of Mn. The energy cutoff for the plane wave basis set was 350 eV. The surface states were calculated with a slab model of four [MBT+BT] layers (48 atoms in total), where the surfaces were fully relaxed under the experimental A-type antiferromagnetic configuration with the presence of DFT-D3 van der Waals correction[44]. A k-mesh of $9 \times 9 \times 1$ was employed. The spin-orbit coupling was considered in all electronic structure calculations except for the surface relaxation.

**Acknowledgment** The authors acknowledge very helpful discussions with Jonathan Sobota and Patrick Kirchmann from SLAC National Accelerator Laboratory, as well as Sarah King, Jiwoong Park, Peter Littlewood, and Supratik Guha from the University of Chicago.

**Funding:** This work was partially supported by the U.S. National Science Foundation via Grant No. DMR-2145373, and by the U.S. Department of Energy (Grant No. DE-SC0022960). The financial support for a part of sample preparation by S.H.L. was provided by the National Science Foundation through the Penn State 2D Crystal Consortium-Materials Innovation Platform (2DCC-MIP) under NSF cooperative agreement DMR-2039351. The sample synthesis efforts by Y.D.G. were supported by the U.S. Department of Energy under grant DE-SC0019068. C.X.L. and R.B.M. acknowledge support from the Penn State MRSEC Center for Nanoscale Science via NSF Grant No. DMR-2011839, and the support of the U.S. Department of Energy (Grant No. DESC0019064). B.Y. acknowledges the financial support by the European Research Council (ERC) under the European Union's Horizon 2020 research and innovation programme (Grant No. 815869).

**Author contributions:** W.L., S.F.-M., C.Y., and S.-L.Y. conducted the trARPES measurements. Y.G., S.H.L., and Z.M. grew the (MnBi$_2$Te$_4$)(Bi$_2$Te$_3$) samples. H.T. and B.Y. performed first-principles calculations with input from R.M. and C.X.L. W.L., S.F.-M., and S.-L.Y. wrote the manuscript with input from all co-authors. S.-L.Y. conceived the experiment.

**Competing interests:** The authors declare that they have no competing interests.

**Data and materials availability:** The data that support the findings of this study are available from the corresponding author upon reasonable request.

## Supplementary Information

Supplementary notes, Figs. S1 to S12

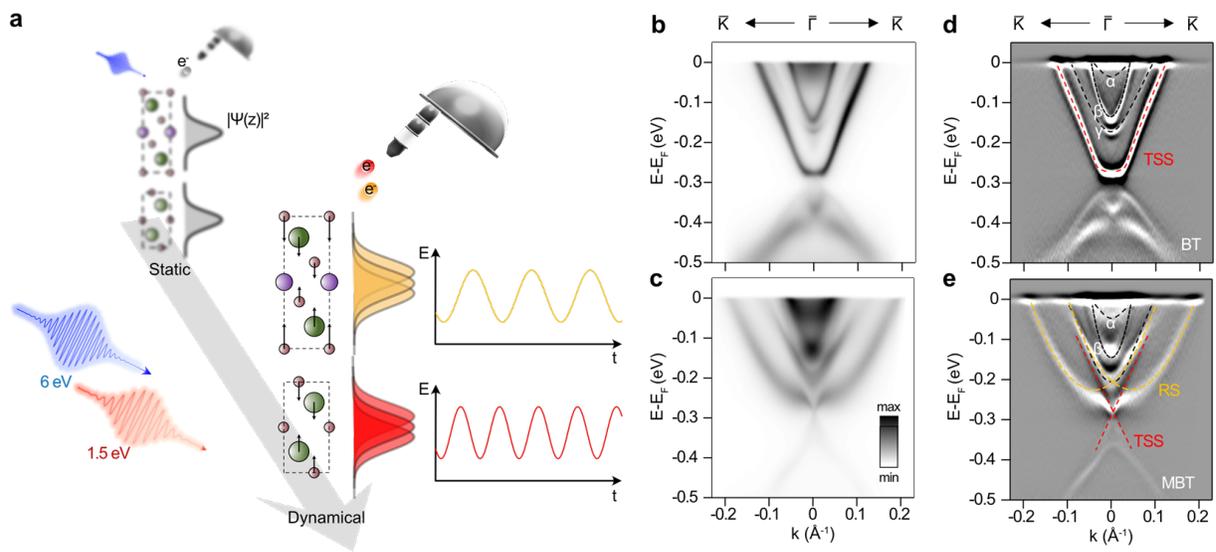

**Fig. 1. Experimental scheme and the electronic band structure of a (MnBi$_2$Te$_4$)(Bi$_2$Te$_3$) superlattice.**
(**a**) Schematics of ARPES and time-resolved ARPES measurements. (**b** and **c**) static micro (μ)-ARPES spectra cutting through $\bar{\Gamma} - \bar{K}$ on (**b**) the Bi$_2$Te$_3$ (BT) termination and (**c**) the MnBi$_2$Te$_4$ (MBT) termination, respectively. (**d** and **e**) Second derivative images of the spectra in (**b**) and (**c**) respectively. The black, red, and yellow dashed lines highlight the bulk bands, topological surface states (TSS) and Rashba-split states (RS), respectively.

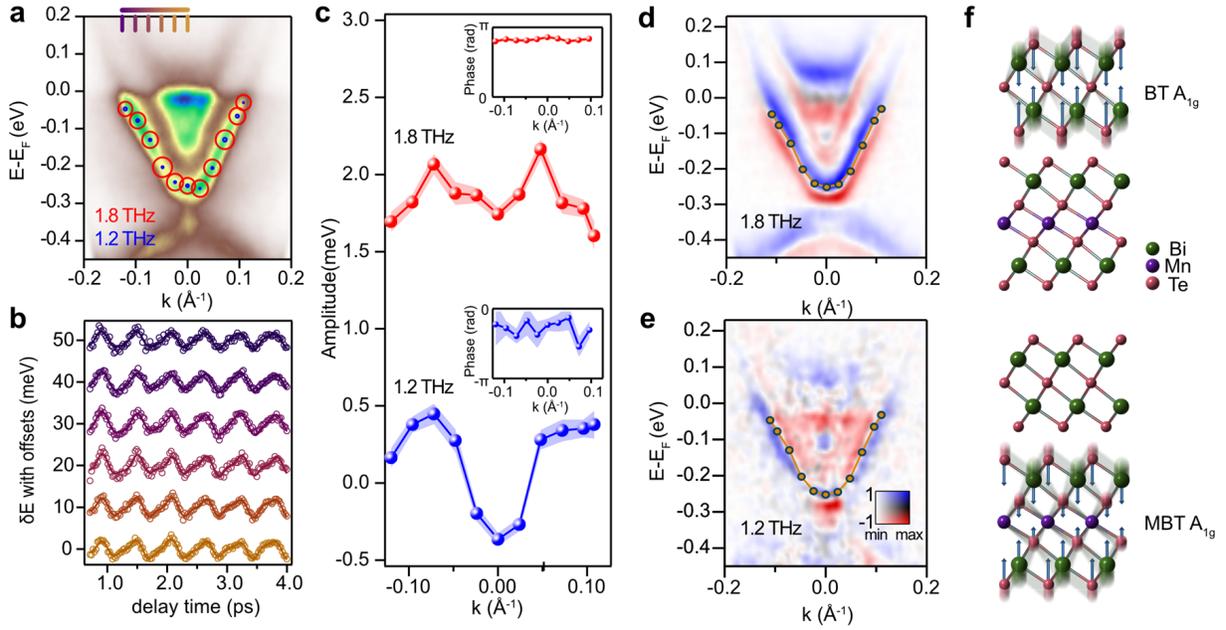

**Fig. 2. Coherent response of the TSS to phonon oscillations on the Bi$_2$Te$_3$ termination.** (**a**) Band dispersions cutting through $\bar{\Gamma} - \bar{K}$ taken at t = 306 fs. The sizes of the red and blue circles represent the oscillation amplitudes of the 1.8 and 1.2 THz modes, respectively. (**b**) Oscillatory components of the TSS energy dynamics taken at different momentum k-points indicated by the gradient color code in (**a**). The oscillations are fitted to cosine functions. (**c**) Oscillation amplitudes of the 1.2 and 1.8 THz modes, with the initial phases shown in the insets. The shaded widths indicate the error bars, which represent one-standard-deviation (1$\sigma$) uncertainties of the fittting results. (**d** and **e**) Phase maps in frequency-domain ARPES (FD-ARPES) at the frequencies of the BT A$_{1g}$ mode (1.8 THz) and the MBT A$_{1g}$ mode (1.2 THz). The red-blue contrast indicates $\cos(\phi)$ where $\phi$ is the initial phase of the oscillation. The intensity saturation indicates the oscillation amplitude of FD-ARPES. In each FD-ARPES map, the intensities are normalized by the maximum value in the energy-momentum space. (**f**) Cartoons of atomic vibrations corresponding to the BT and MBT A$_{1g}$ modes.

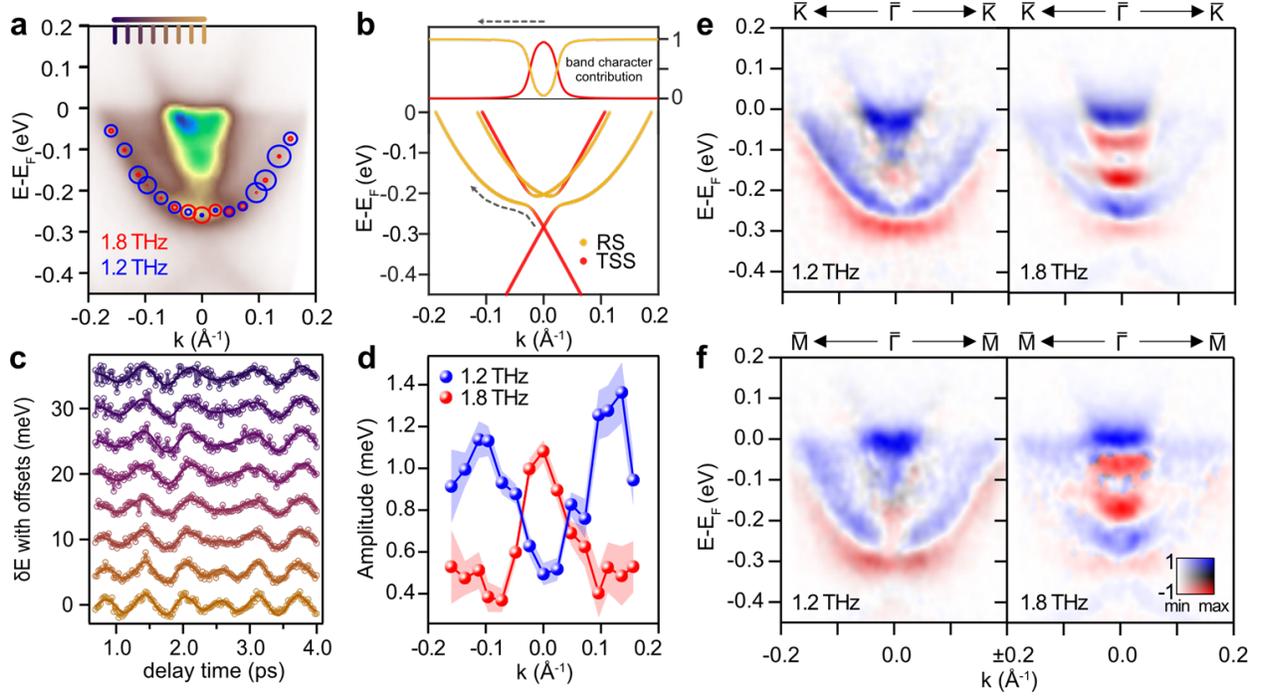

**Fig. 3. Coherent response on the MnBi$_2$Te$_4$ termination.** (a) trARPES spectra cutting through $\bar{\Gamma} - \bar{K}$ taken at t = 306 fs. The sizes of the red and blue circles represent the oscillation amplitudes of the 1.8 and 1.2 THz modes, respectively. (b) Simulation of the RS-TSS hybridization using an effective Hamiltonian (Supplementary Note 1). The band characters on the modified RS are also plotted. (c) Oscillatory components of the RS energy dynamics taken at different momentum k-points indicated by a gradient color code in (a). (d) Oscillation amplitudes of the 1.2 and 1.8 THz modes, extracted by fitting coherent energy oscillations to two-cosine functions. The shaded widths indicate the error bars, which represent one-standard-deviation (1$\sigma$) uncertainties of the fittting results. (e and f) Frequency-domain ARPES (FD-ARPES) spectra of the 1.2 and 1.8 THz modes on the MnBi$_2$Te$_4$ termination along (e) the $\bar{\Gamma} - \bar{K}$ dispersion and along (f) the $\bar{\Gamma} - \bar{M}$ dispersion. The red-blue contrast indicates $\cos(\phi)$ where $\phi$ is the initial phase of the oscillation. The intensity saturation indicates the oscillation amplitude of FD-ARPES. In each FD-ARPES map, the intensities are normalized by the maximum value in the energy-momentum space.

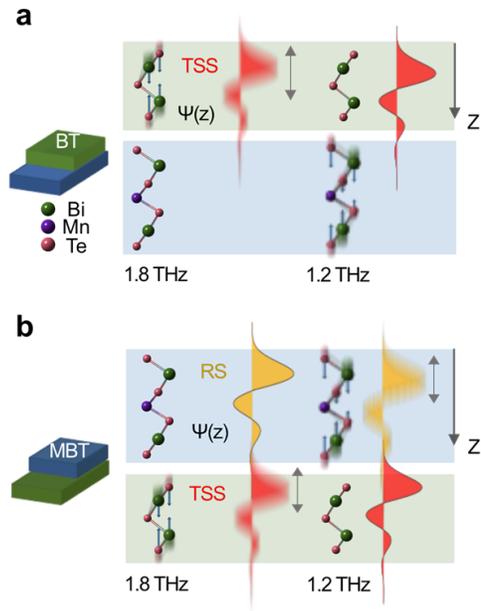

**Fig 4. Schemes of layer-by-layer disentanglement of Bloch states.** (**a** and **b**) The BT and MBT $A_{1g}$ phonon modes, as well as cartoons of the TSS and RS wavefunctions on (**a**) the BT termination and (**b**) the MBT termination.